\documentclass[twocolumn,showpacs,preprintnumbers,amsmath,amssymb]{revtex4}

\usepackage{graphicx}
\usepackage{dcolumn}
\usepackage{bm}

\begin{document}

\title{Viscous flow over a chemically patterned surface}

\author{J.E. Sprittles}
\email{sprittlj@maths.bham.ac.uk}

\author{Y.D. Shikhmurzaev}
\email{Y.D.Shikhmurzaev@bham.ac.uk}


\affiliation{School of Mathematics, University of Birmingham,
Birmingham,  B15 2TT, UK.}

\date{\today}

\begin{abstract}
The classical fluid dynamics boundary condition of no-slip
suggests that variation in the wettability of a solid should not
affect the flow of an adjacent liquid.  However experiments and
molecular dynamics simulations indicate that this is not the case.
In this paper we show how flow over a solid substrate with
variations of wettability can be described in a continuum
framework using the interface formation theory developed earlier.
Results demonstrate that a shear flow over a perfectly flat solid
surface is disturbed by a change in its wettability, i.e.\ by a
change in the chemistry of the solid substrate. The magnitude of
the effect is shown to be proportional to
$\cos\theta_1-\cos\theta_2$ where $\theta_1$ and $\theta_2$ are
the equilibrium contact angles that a liquid-gas free surface
would form with the two chemically different parts of the solid
surface.
\end{abstract}


\maketitle

\section{Introduction}

The flow of liquids over chemically patterned surfaces is an
exciting and relatively new area of fluid mechanics with
applications in many emerging technologies \cite{xia01}.  Such
flows are of particular interest in microfluidics where an
increasing surface to volume ratio of liquids means that surface
effects become of greater significance \cite{darhuber05}. The
correct description of the physics at solid-liquid interfaces then
becomes imperative to the success of any attempt to model this
class of flows \cite{quake05}.  It has been shown that by
patterning a substrate with hydrophillic and hydrophobic regions
it is possible to confine a liquid to a microchannel \cite{zhao01,
gau99}, to improve the accuracy of droplet deposition
\cite{burns03, dupuis04} or to create a structured film
\cite{braun99}. Alternatively, when a wettability gradient is
present, unbalanced surface tension forces can lead to the
movement of liquid drops \cite{chaudhury92, daniel01}.

The effect of variable wettability of the solid substrate on the
adjacent flow has been studied theoretically using molecular
dynamics simulations \cite{priezjev05, qian05}.  In this approach,
variations in wettability are modelled by varying an interaction
potential between molecules of the solid and the fluid.  The
results show that a change in wettability does affect the flow,
most notably producing a component of velocity normal to the solid
surface. Given that ``wettability" can be introduced as a
macroscopic characteristic of a liquid-solid system, one should be
able to model the effects discovered by molecular dynamics
macroscopically using an appropriate formulation in the framework
of continuum mechanics.

In the continuum approximation, molecular properties of the
contacting media, as well as molecular length and time scales, do
not appear explicitly.  Instead, they manifest themselves in the
bulk equations and boundary conditions via macroscopic transport
coefficients and parameters of constitutive equations.  The
classical no-slip boundary condition leaves no room for
incorporating the effects of variable wettability and, as is well
known \cite{dussan74, dussan79}, it becomes inadequate for dealing
with processes of dynamic wetting where wettability of the solid
plays a key role.

A generalisation of no-slip often cited in the literature is the
Navier slip condition \cite{Navier23, Lamb32}, where slip, i.e.\
the difference between the tangential velocities of the fluid and
the solid, is assumed to be proportional to the tangential stress
acting from the fluid on the liquid-solid interface.  The
coefficient of proportionality is referred to as the `slip
coefficient' or the `coefficient of sliding friction'. Various
expressions for this coefficient have been proposed \cite{lauga05}
some of which are based on the results of molecular dynamics
simulations \cite{thompson97}. An appropriate choice of the slip
coefficient allows one to describe situations in which the
characteristic length scales of the flow are small (e.g.
nanopores), so that the effects of slip become apparent
\cite{sokhan01}.

The problem arises when the approach based on the Navier slip
condition is used to describe flows over solid surfaces with
variable wettability, since the existing models provide no
conceptual link between slip and wettability.  Indeed, as reviewed
in \S 9 of \cite{shik97}, in the area of dynamic wetting, where
the physics of wettability becomes a key factor, the so-called
`slip models' treat the behaviour of the contact angle, which is
the macroscopic `measure' of wettability, and slip, which is used
as a boundary condition removing the stress singularity at the
moving contact line, as completely independent. On a macroscopic
level, any coupling of a formula for the velocity-dependence of
the dynamic contact angle and a slip boundary condition will
produce a new slip model.  Alternatively, linking the slip
coefficient (or any other slip condition) to wettability via
molecular dynamics simulations negates the advantages of the
self-contained and experimentally verifiable continuum modelling.

Applying the no-slip or constant slip coefficient Navier boundary
condition at a solid surface implies that the flow of a liquid
will not be affected by variations in the wettability of an
adjacent solid. This is in direct conflict with the conclusions of
molecular dynamics simulations \cite{priezjev05, qian05} that were
specifically tailored to investigate this problem. In the present
paper we consider flow over a solid surface of variable
wettability using a model developed earlier \cite{shik93} that
incorporates wettability in the framework of continuum mechanics
from the following perspective. Given that dynamic wetting is, by
its very name, a process of creating a new/fresh liquid-solid
interface (`a wetted solid surface'), an adequate mathematical
model of dynamic wetting must have the physics of interface
formation process at its core. The model (hereafter referred to as
the interface formation model) has been derived from first
principles \cite{shik93} using methods of irreversible
thermodynamics and notably it ensures that all aspects of
wettability (contact angle, slip, adsorption) are interrelated.
Embedding dynamic wetting in a general physical framework, as a
particular case from a broad class of flows with
forming/disappearing interfaces, allows one to use the same model,
without any ad hoc alterations, to describe other flows from this
class \cite{shik05a, shik05b, shik05c}. This makes it possible to
use the values for material constants determined from completely
independent experiments.  Comparison with such experiments have
confirmed the model's validity and accuracy \cite{shik93,
blake02}.

We will examine the fully representative case of a plane-parallel
shear flow over a smooth solid surface that encounters a
transition region between substrates of different wettability.

\section{Problem formulation}

Consider the steady flow of an incompressible Newtonian liquid
passing over a stationary flat solid surface. The liquid is driven
over the solid by a plane-parallel shear of magnitude $S$ in the
far field. We consider the bulk flow to be described by the
Navier-Stokes equations
\begin{equation}\label{incom} \nabla\cdot\mathbf{u}=0,~~~~ \rho \mathbf{u} \cdot \nabla \mathbf{u} =-\nabla p+\mu\nabla^{2}
\mathbf{u}
\end{equation}
where $\mathbf{u}$, $p$, $\rho$ and $\mu$ are the fluid's
velocity, pressure, density and viscosity, respectively. The
boundary conditions on a solid surface of uniform wettability,
that follow from the interface formation theory \cite{shik93} are
given by
\begin{equation}\label{gen nav}
\mu\mathbf{n}\mathbf{\cdot[\nabla u +
(\nabla u)^{*}]}\cdot(\mathbf{I-n}\mathbf{n}) +
\hbox{$\frac{1}{2}$}\nabla \sigma=\beta
\mathbf{u}\cdot(\mathbf{I-n}\mathbf{n}),
\end{equation}
\begin{equation}\label{norm flux}
\rho\mathbf{u\cdot n} =
\frac{\rho^{s}-\rho^{s}_{e}}{\tau},
\end{equation}
\begin{equation}\label{cont dens}
\nabla\cdot(\rho^{s}\mathbf{v}^{s})=
-\frac{\rho^{s}-\rho^{s}_{e}}{\tau},
\end{equation}
\begin{equation}\label{chan flow}
\left( \mathbf{{v}}^{s}-\hbox{$\frac{1}{2}$}
\mathbf{u}\right)\cdot(\mathbf{I-n}\mathbf{n})=\alpha\nabla
\sigma,
\end{equation}
\begin{equation}\label{state} \sigma=\gamma(\rho^{s}_{(0)}-\rho^{s}). \end{equation}
Here $\sigma$ is the surface tension in the `surface phase', i.e.
a microscopic layer of liquid adjacent to the solid surface
subject to intermolecular forces from two bulk phases; $\rho^{s}$
is the surface density (mass per unit area) of this layer and
$\mathbf{v}^{s}$ is the velocity with which it is transported;
$\mathbf{n}$ is the unit vector normal to the solid surface
pointing into the liquid; $\mathbf{I}$ is the metric tensor;
$(\mathbf{I-n}\mathbf{n})$ is a tensor that extracts the
tangential component of a vector, for example
$\mathbf{u}\cdot(\mathbf{I-n}\mathbf{n})=\mathbf{u}_{||}$, where
the double line subscript denotes tangential component; $\alpha$,
$\beta$, $\gamma$, $\tau$, $\rho_{(0)}^{s}$ and $\rho_{e}^{s}$ are
phenomenological material constants.

The model has previously been discussed in detail (e.g.
\cite{shik93}), so that here we will only briefly recapitulate the
meaning of the terms. The surface tension is considered as a
dynamic quantity related to the surface density via the equation
of state in the `surface phase' ($\ref{state}$) which is taken
here in the simplest linear form. The constant $\gamma$ is
associated with the inverse compressibility of the fluid whilst
$\rho_{(0)}^{s}$ is the surface density corresponding to zero
surface tension. As in Gibbs' theory of capillarity, the surface
tension itself may be positive or negative depending on whether
the solid is hydrophobic or hydrophillic. Gradients in surface
tension influence the flow, firstly, via the tangential stress
boundary condition ($\ref{gen nav}$), i.e. via the Marangoni
effect, and, secondly, in (\ref{chan flow}) by forcing the surface
velocity $\mathbf{v^{s}}$, which is parallel to the solid surface,
to deviate from that generated in the surface phase by the outer
flow. The constants $\alpha$ and $\beta$ characterise the response
of the interface to surface tension gradients and an external
torque, respectively; in the simplest variant of the theory both
are properties of the fluid and have no relation to the
wettability of the solid. Mass exchange between the bulk and
surface phases, caused by the possible deviation of the surface
density from its equilibrium value $\rho^{s}_{e}$,  is accounted
for in the boundary condition for the normal component of bulk
velocity ($\ref{norm flux}$) and in the surface mass balance
equation (\ref{cont dens}). The parameter $\tau$ is the surface
tension relaxation time.

One would expect a generalised set of boundary conditions to have
the no-slip condition as its limiting case. For the interface
formation model this limiting case follows from the limits
$\mu/(\beta L) ,~U\tau/L \rightarrow 0$, where $L$ and $U$ are
characteristic length and velocity scales of the flow, applied to
(\ref{gen nav})--(\ref{state}). Estimates for the phenomenological
constants $\beta$ and $\tau$, obtained from experiments on dynamic
wetting \cite{blake02}, suggest that, say, for $L\sim 10^{-2}$~cm,
$U\sim 1$~cm~s$^{-1}$ and $\mu\sim 10$~g~cm$^{-1}$~s$^{-1}$ one
has $\mu/(\beta L)\sim10^{-5}$ and $U\tau/L\sim10^{-6}$.  Then for
flows which, unlike dynamic wetting, are not associated with
infinitesimal length scales, boundary conditions (\ref{gen
nav})--(\ref{state}) to leading order in $\mu/(\beta L)$ and
$U\tau/L$, reduce to no-slip.

Conditions (\ref{gen nav})--(\ref{state}) were derived using
methods of irreversible thermodynamics assuming that the
equilibrium surface density $\rho^{s}_{e}$, determined by the
wettability of the solid substrate, is a constant, i.e. for a
solid surface of a uniform wettability. This constant is related
to the equilibrium contact angle that a liquid-gas free surface
would form with the solid via the Young equation in the following
way. If we assume that the equilibrium surface tension of the
solid-gas interface is negligible, the Young equation takes the
form
\begin{equation}\label{young}
\sigma_{sl}=-\sigma_{lg}\cos\theta,
\end{equation}
where $\sigma_{sl}$ and $\sigma_{lg}$ are the equilibrium surface
tensions of the solid-liquid and liquid-gas interfaces and
$\theta$ is the equilibrium contact angle that the free surface
forms with the solid.  Thus, since according to (\ref{state}),
$\sigma_{sl}=\gamma(\rho^{s}_{(0)}-\rho^{s}_{e})$ for a given
liquid the Young equation (\ref{young}) allows one to express
$\rho^{s}_{e}$ in terms of $\theta$ (see (\ref{rhotheta}) below).

In order to modify ($\ref{gen nav}$)--($\ref{state}$) for a solid
of variable wettability we must replace, in ($\ref{gen nav}$) and
($\ref{chan flow}$)
\begin{equation}\label{new} \nabla \sigma \rightarrow \nabla
\sigma+ \rho^{s} F^{s},~~\hbox{where}~~ F^{s}=\frac{\gamma\nabla
\rho^{s}_{e}}{\rho^{s}_{e}}.
\end{equation}
The reaction force $F^{s}$ acts on the liquid-solid interfacial
layer from the solid surface and by balancing gradients of the
equilibrium surface tension
$\sigma_{e}=\sigma\left(\rho^{s}_{e}\right)$, ensures the
existence of a state of equilibrium.

Consider the equilibrium surface density of the form
\begin{equation}\label{rhose}
\rho^{s}_{e}=\hbox{$\frac{1}{2}$}\left(\rho^{s}_{1e}+\rho^{s}_{2e}\right)+
\hbox{$\frac{1}{2}$}\left(\rho^{s}_{2e}-\rho^{s}_{1e}\right)\tanh\left(x/l\right),
\end{equation}
where $\rho^{s}_{1e}$ and $\rho^{s}_{2e}$ are the equilibrium
surface densities in the far field as $x\rightarrow-\infty$ and
$x\rightarrow\infty$, respectively; $l$ is the length of the
transition region; $x$ is a Cartesian coordinate in the plane of
the solid surface.  The equilibrium surface density in the form of
(\ref{rhose}) allows one to investigate the role of a transition
region between solids of different wettabilities and, after taking
the limit $l/L\rightarrow 0$, find what boundary conditions for
the surface variables one should use if this region is modelled as
a solid-solid-liquid contact line.

Finally, we will assume that the flow is plane-parallel in the
$(x,y)$-plane of a Cartesian coordinate system, the origin of
which is at the centre of the transition region, and that it is
generated by constant shear of magnitude $S$ in the far field,
\begin{equation}\label{far field} \mathbf{u}\rightarrow S~\mathbf{i
j}\cdot\mathbf{r}~~\hbox{as}~~\mathbf{r}\rightarrow\infty,
\end{equation}
where $\mathbf{i}$ and $\mathbf{j}$ are unit vectors in the $x$
and $y$ directions and $\mathbf{r}$ is the radius-vector.
Equations (\ref{incom})--(\ref{far field}) now completely specify
the problem.

It is important to emphasise that in the derivation of the model
it is assumed that on the solid-facing side of the solid-liquid
interface one has impermeability and no-slip.  However, it is the
velocity on the {\it liquid}-facing side of the solid-liquid
interface that is the boundary condition for the Navier-Stokes
equations (\ref{incom}).  In the classical no-slip condition it is
assumed that there is no difference in velocity between the
solid-facing and liquid-facing side of the interface, whereas in
the interface formation model, the velocity on the liquid-facing
side of the interface is determined by the interaction occurring
in the surface phase and between the surface phase and the bulk.
As a result, one can expect effective (or `apparent') slip, i.e.\
the difference between the velocity on the liquid-facing side of
the interface and the velocity of the solid surface, that is, in
our case, $\mathbf{u}\cdot(\mathbf{I-n}\mathbf{n})\neq0$) and a
non-zero normal component of velocity (a flux in/out of the
surface phase, i.e.\ $\mathbf{u}\cdot\mathbf{n}\neq0$).

It is convenient to non-dimensionalise equations
(\ref{incom})--(\ref{far field}) using
\begin{eqnarray*}\label{nd quant} U=\mu^{-1}\sigma_{lg},\quad  L=US^{-1},
\quad P=\mu S,\quad \sigma_{lg},\quad
\rho^{s}_{(0)}\end{eqnarray*}
as the scales for velocities, length, pressure, surface tension
and surface density. Then in the bulk one has
\begin{equation}\label{incomnd}
\nabla\cdot\mathbf{u}=0,~~~~ Re\left(\mathbf{u} \cdot \nabla \mathbf{u}\right) =-\nabla p+\nabla^{2}
\mathbf{u}.
\end{equation}
whilst on the surface, where we use the notation $u$ and $v$ for
tangential and normal components of velocity,
\begin{equation}\label{gen navnd}
\left(\frac{\partial u}{\partial y} +
\frac{\partial v}{\partial x}\right) + \frac{1}{2}\left(\frac{d
\sigma}{d x}+\frac{\lambda\rho^{s}}{\rho^{s}_{e}}\frac{d
\rho^{s}_{e}}{d x}\right)=\bar{\beta}u,
\end{equation}
\begin{equation}\label{norm fluxnd}  v =
Q\left(\rho^{s}-\rho^{s}_{e}\right),
\end{equation}
\begin{equation}\label{cont densnd} \epsilon\frac{ d\left(\rho^{s}v^{s}\right)}{d x}=
-\left(\rho^{s}-\rho^{s}_{e}\right),
\end{equation}
\begin{equation}\label{chan flownd}  v^{s}=\hbox{$\frac{1}{2}$}u
+\bar{\alpha}\left(\frac{d \sigma}{d x}
+\frac{\lambda\rho^{s}}{\rho^{s}_{e}} \frac{d \rho^{s}_{e}}{d
x}\right),\end{equation}
\begin{equation}\label{statend} \sigma=\lambda(1-\rho^{s}), \end{equation}
\begin{equation}\label{ca rhose} \rho^{s}_{e}=\hbox{$\frac{1}{2}$}\left(\bar{\rho}^{s}_{1e}+\bar{\rho}^{s}_{2e}\right)+
\hbox{$\frac{1}{2}$}\left(\bar{\rho}^{s}_{2e}-\bar{\rho}^{s}_{1e}\right)\tanh\left(x/\bar{l}\right),
\end{equation}
and in the far field
\begin{equation}\label{far fieldnd}
u\rightarrow 1~,v\rightarrow 0 ~~\hbox{as}~~x^{2}+y^{2}\rightarrow\infty.
\end{equation}
Here

$$
Re=\frac{\rho \sigma_{lg}^{2}}{S\mu^{3}},\ \epsilon=S\tau,\
\bar{\beta}=\frac{\beta\sigma_{lg}}{\mu^{2}S},\
Q=\frac{\rho_{(0)}^{s}\mu}{\rho\tau\sigma_{lg}},\
\bar{\alpha}=\frac{\alpha S \mu^{2}}{\sigma_{lg}},
$$
$$
\bar{l}=\frac{\mu S l}{\sigma_{lg}},\
\lambda=\frac{\gamma\rho^{s}_{(0)}}{\sigma_{lg}},\
\bar{\rho}^s_{ie}=\frac{\rho^s_{ie}}{\rho^{s}_{(0)}}\qquad(i=1,2).
$$

It is noteworthy that, unlike the classical Navier condition, in
(\ref{gen navnd}), which can be regarded as its generalisation,
the tangential stress (the first term on the left-hand side)
includes $\partial v/\partial x$. The deviation of $\rho^{s}$ from
its equilibrium value and the resulting adsorption/desorption lead
to $v\neq0$ via (\ref{norm fluxnd}). Then, the spatial
non-uniformity of this process makes $\partial v/\partial x\neq0$
thus `switching on' this term in the tangential stress in
(\ref{gen navnd}) and hence contributing to the apparent slip.

The parameter $Q$ in (\ref{norm fluxnd}) is the ratio of the
characteristic mass flux into (out of) the liquid-solid interface
associated with the adsorption (desorption) process triggered by
the deviation of the surface density from its local equilibrium
value and the characteristic mass flux due to convection in the
bulk, whereas $\epsilon^{-1}$ characterizes the ratio of the
former and the characteristic {\it divergence\/} of the convective
mass flux in the surface phase. In the present context, $\epsilon$
is simply the product of the magnitude of shear in the far field
$S$ and the surface-tension-relaxation time $\tau$.

Given that it is the equilibrium contact angle $\theta$, which a
liquid-gas free surface would form with a solid, that we use as a
measure of the wettability of a solid substrate, it is convenient
to eliminate $\bar{\rho}^{s}_{ie}$, which are `internal'
parameters of the model, in favour of $\theta_{i}$ using
(\ref{young}) and (\ref{statend}):
\begin{equation}\label{rhotheta}
\bar{\rho}^{s}_{ie}=1+\lambda^{-1}\cos\theta_{i}\qquad(i=1,2).
\end{equation}

Hereafter we will refer to the portion of the solid substrate with
equilibrium contact angle $\theta_{1}$ and $\theta_{2}$ as `solid
$1$' and `solid $2$', respectively.

Following from \cite{blake02}, where it was shown by analysing
experiments on dynamic wetting that $\alpha\sim\beta^{-1}$, we
will further assume that $\alpha\beta=1$ and hence that, in terms
of our non-dimensional parameters, $\bar{\alpha}\bar{\beta}=1$.
The analysis of experiments in \cite{blake02} also provides
estimates for the magnitude of phenomenological constants in the
interface formation model's equations.  Using these estimates and
taking $\rho\sim 1$ g~cm$^{-3}$, $\mu\sim 10^{-1}-10^{2}$
g~cm$^{-1}$~s$^{-1}$, $\sigma_{lg} \sim 10-10^{2}$ dyn~cm$^{-1}$,
$S\sim 10^{3}-10^{5}$ s$^{-1}$ and $l\sim 10^{-6}-10^{-5}$ cm one
arrives at a typical range of values for the magnitudes of the
non-dimensional groups
\begin{align*}
Re&\sim 10^{-9}-10^{3},
&\epsilon&\sim 10^{-6}-10^{-2},%
&\bar\beta&\sim 10^{1}-10^{7},\\
Q&\sim 10^{-3}-10^{3},%
&\bar{l}&\sim 10^{-6}-10^{1}, %
&\lambda&\sim 2-10^{2},\\
\theta_{i}&\in [0^{\circ},180^{\circ}] &(&i=1,2).
\end{align*}
As mentioned earlier, it is the double limit
$\bar{\beta}^{-1}\rightarrow0, ~\epsilon\rightarrow0$ applied to
equations (\ref{gen navnd})--(\ref{statend}) that results in the
no-slip condition and hence all effects associated with deviation
from the classical no-slip are at leading order in these
parameters.

\section{Solution}
\begin{figure}[h]
\centering
\includegraphics[scale=0.5]{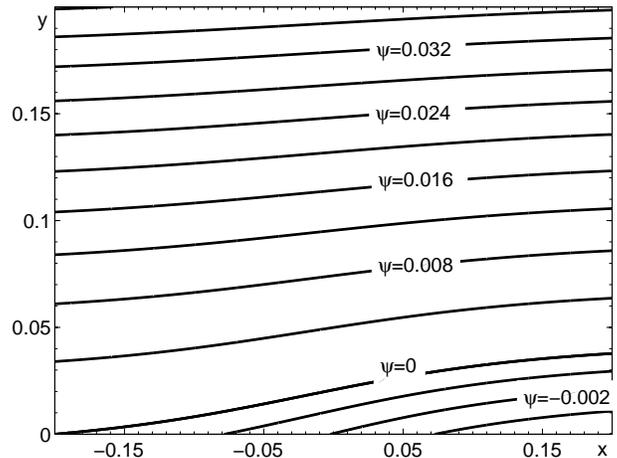}
\caption{Streamlines of flow over a chemically patterned surface
in which values of the streamfunction $\psi$ are given. Parameter
values $Re=0.01$, $\epsilon=0.01$, $\bar{\beta}=10$, $Q=100$,
$\bar{l}=0.1$, $\lambda=10$, $\theta_{1}=10^{\circ}$ and
$\theta_{2}=100^{\circ}$ .} \label{F:stream}
\end{figure}

The problem was solved numerically using the finite element
method.  Fig.~\ref{F:stream} shows the streamlines of the flow for
the case where solid $1$ is more hydrophillic than solid $2$; the
values of the dimensionless constants are given in the figure
caption.  As one can see, when the outer flow drives the fluid
from a hydrophillic to a hydrophobic zone, there appears a normal
flux from the surface phase into the bulk. It is noteworthy that,
as shown in Fig.~\ref{F:stream}, the vertical component of bulk
velocity is nonzero at $y=0$, whereas for the classical Navier
condition with different coefficients of sliding friction one has
$v=0$ at $y=0$ \cite{qian05} and the normal component of velocity
away from the solid appears solely due to the disturbance of the
tangential flow at $y=0$.

\begin{figure}
\centering
\includegraphics[scale=0.3]{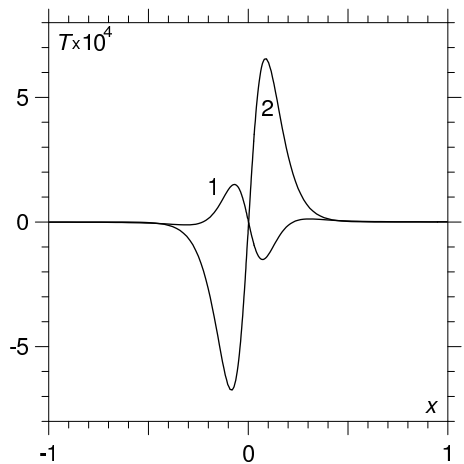}
\caption{Variation of the `generators' of slip (the terms on the
left hand side of the generalised Navier condition (\ref{gen
navnd})) in response to a variation in the solid surface
wettability. In curve $1$, T=$\left[\partial u/\partial y+\partial
v/\partial x-1\right]$ whilst in curve $2$, T=$\left[d \sigma/d
x+\left(\lambda\rho^{s}/\rho^{s}_{e}\right) d \rho^{s}_{e}/d
x\right]/2$. Results are obtained for parameter values $Re=0.01$,
$\epsilon=0.01$, $\bar{\beta}=100$, $Q=1$, $\bar{l}=0.1$,
$\lambda=20$, $\theta_{1}=10^{\circ}$, $\theta_{2}=100^{\circ}$.}
\label{F:gen nav}
\end{figure}

\begin{figure}
\centering
\includegraphics[scale=0.3]{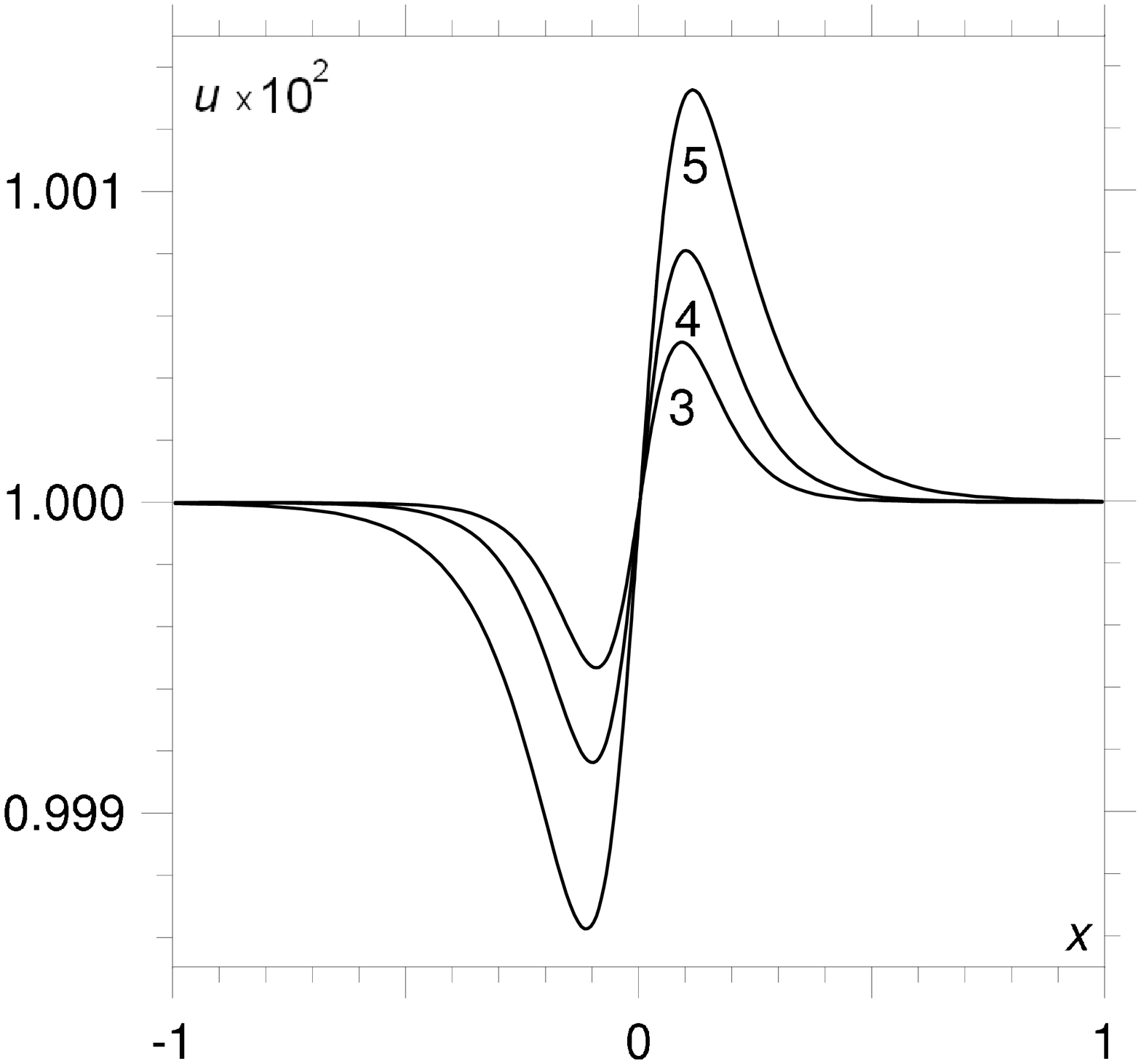}
\includegraphics[scale=0.3]{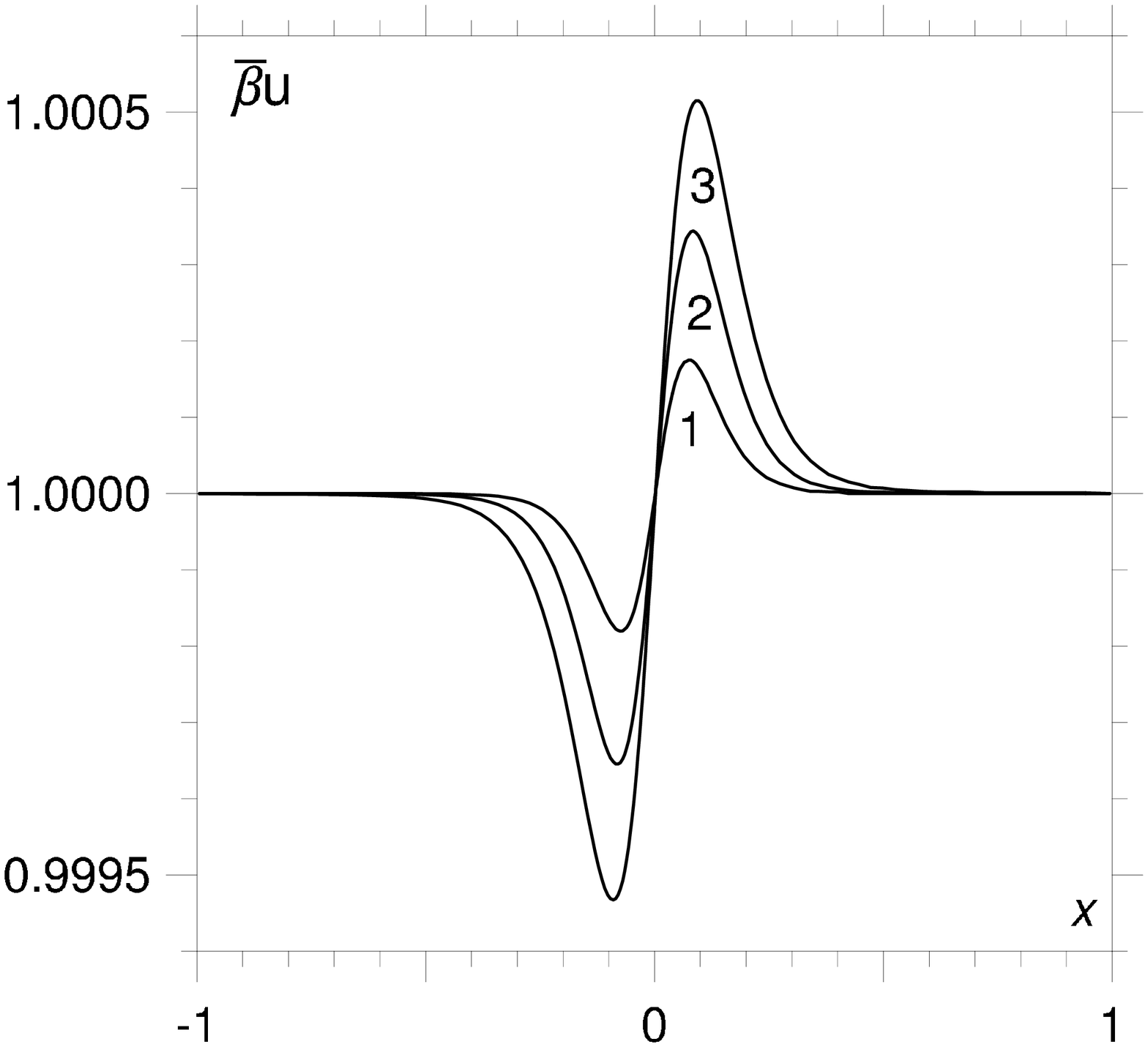}
\includegraphics[scale=0.3]{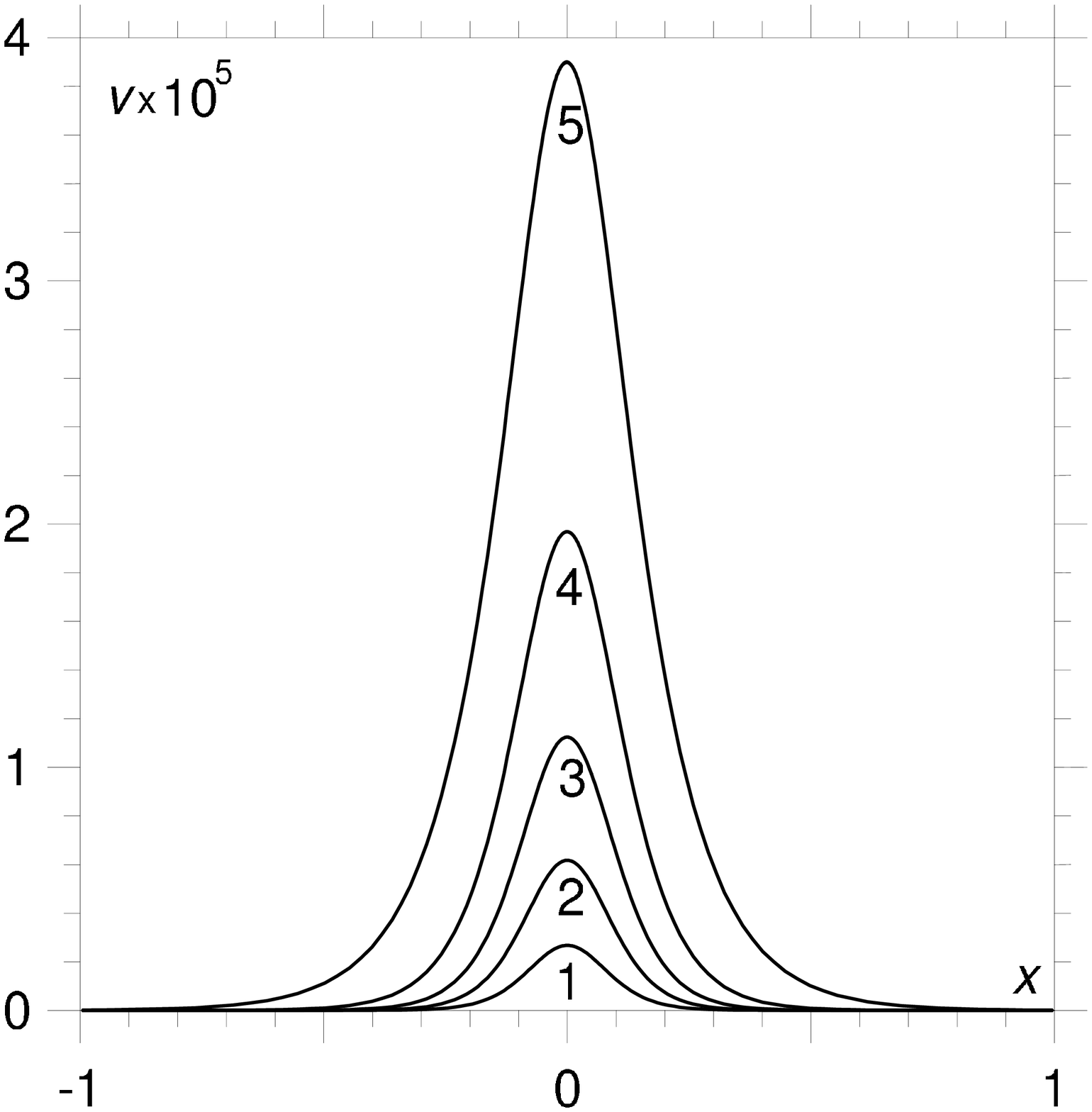}
\caption{Distribution of $u(x,0)$ and $v(x,0)$ for different
$\epsilon$ and $\bar{\beta}$.  Curves $1$--$5$ correspond to
$\epsilon=0.01$, $\bar{\beta}=500$; $\epsilon=0.01$,
$\bar{\beta}=200$; $\epsilon=0.01$, $\bar{\beta}=100$;
$\epsilon=0.02$, $\bar{\beta}=100$; $\epsilon=0.05$,
$\bar{\beta}=100$, respectively; for all curves $Re=0.01$, $Q=1$,
$\bar{l}=0.1$, $\lambda=20$, $\theta_{1}=10^{\circ}$ and
$\theta_{2}=100^{\circ}$.} \label{F:eps_beta}
\end{figure}

\begin{figure}
\centering
\includegraphics[scale=0.3]{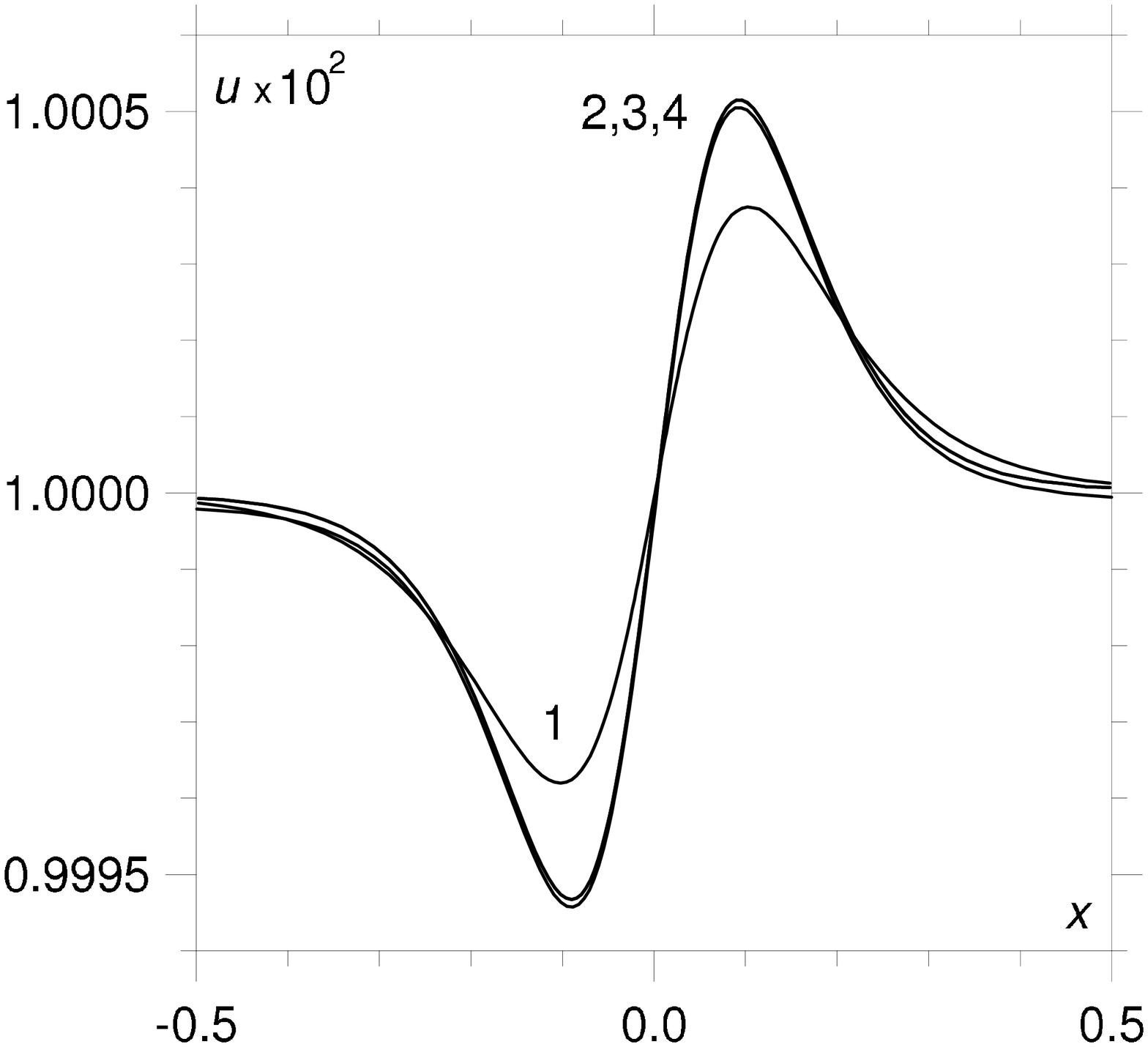}
\includegraphics[scale=0.3]{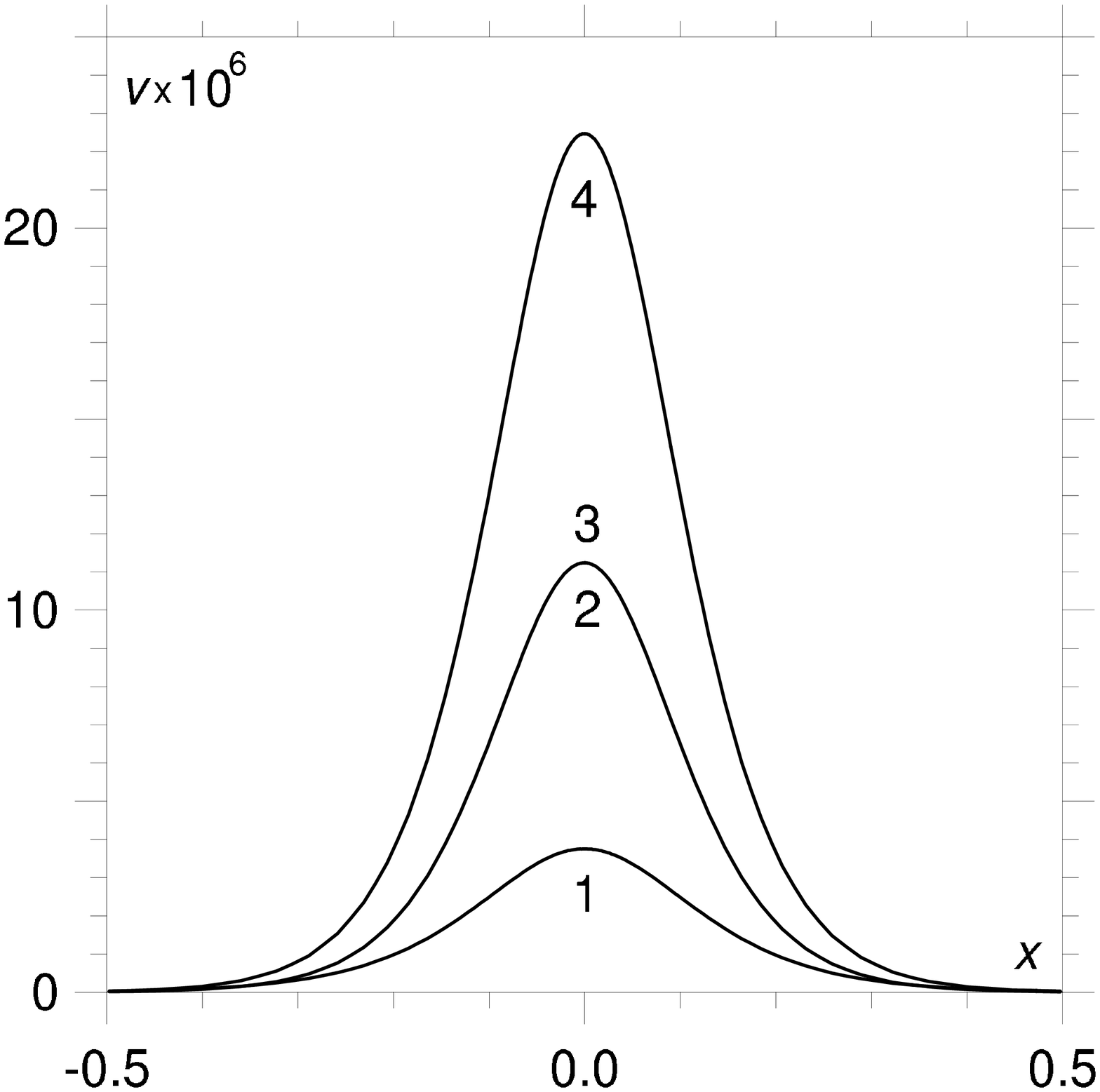}
\caption{Dependence of $u(x,0)$ and $v(x,0)$ on $Re$, $Q$ and
$\lambda$. Curve $2$ corresponds to $Re=0.01$, $Q=1$,
$\lambda=20$; 1: $\lambda=50$; 3 (indistinguishable within the
graphical accuracy): $Re=10$; 4: $Q=2$. Parameters
$\epsilon=0.01$, $\bar{\beta}=100$, $\bar{l}=0.1$,
$\theta_{1}=10^{\circ}$, $\theta_{2}=100^{\circ}$ for all curves.
} \label{F:re_q_la}
\end{figure}

\begin{figure}
\centering
\includegraphics[scale=0.3]{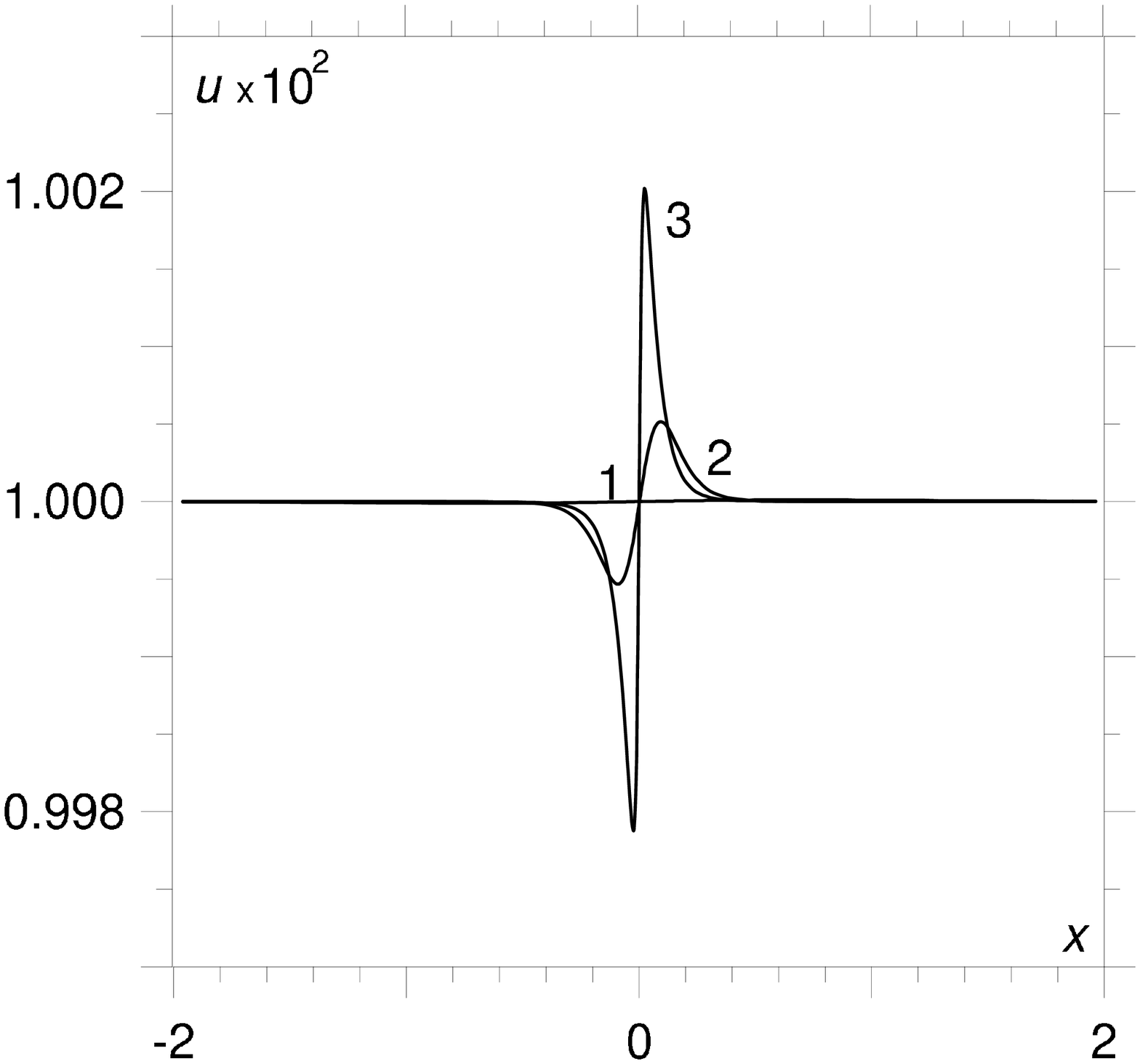}
\includegraphics[scale=0.3]{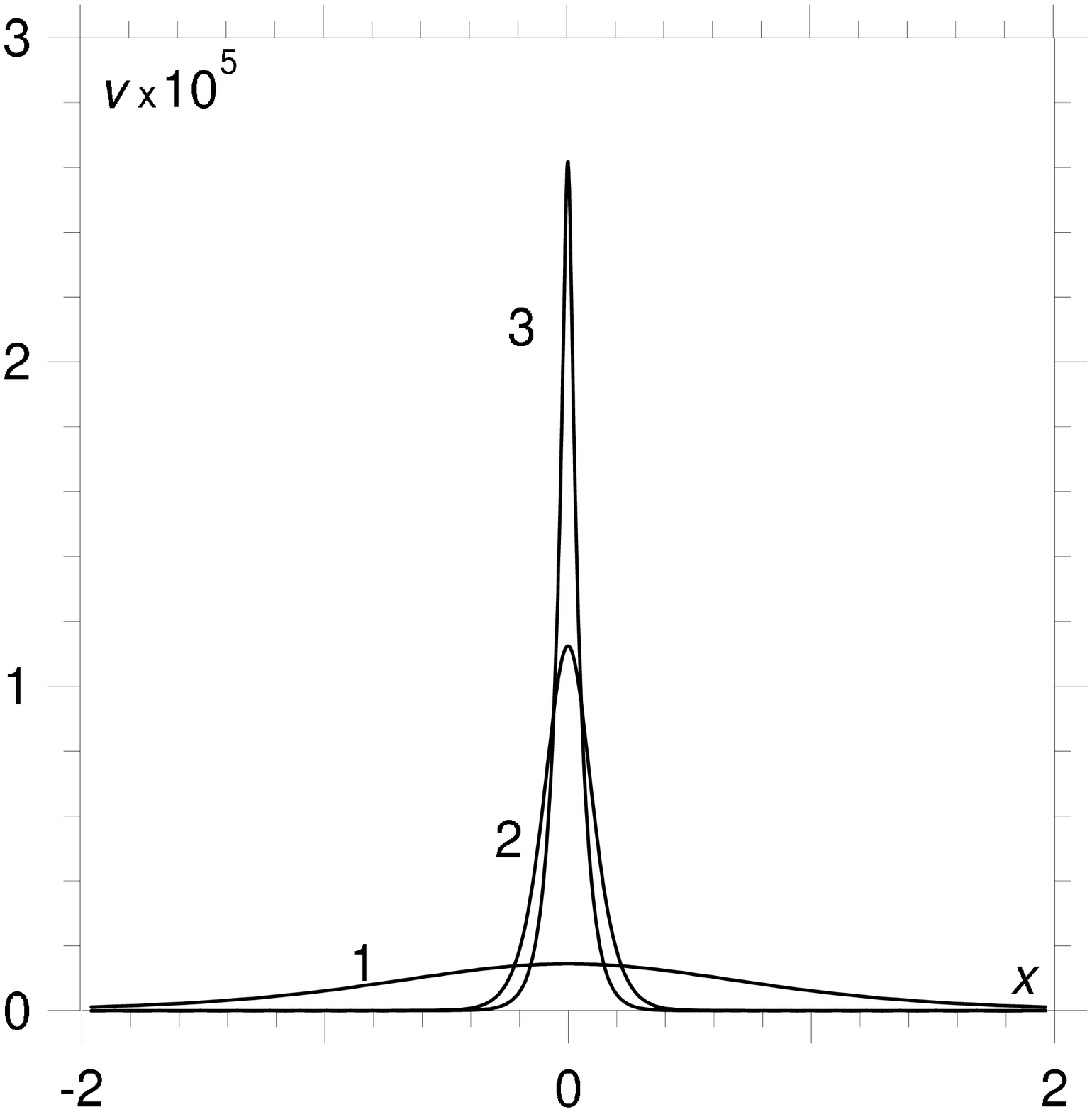}
\caption{Velocity on the surface for different widths of
transition region.  Curves $1$--$3$ correspond to $\bar{l}=1$,
$\bar{l}=0.1$ and $\bar{l}=0.01$, respectively; for all curves
$Re=0.01$, $\epsilon=0.01$, $\bar{\beta}=100$, $Q=1$,
$\lambda=20$, $\theta_{1}=10^{\circ}$ and
$\theta_{2}=100^{\circ}$. } \label{F:l}
\end{figure}

\begin{figure}
\centering
\includegraphics[scale=0.3]{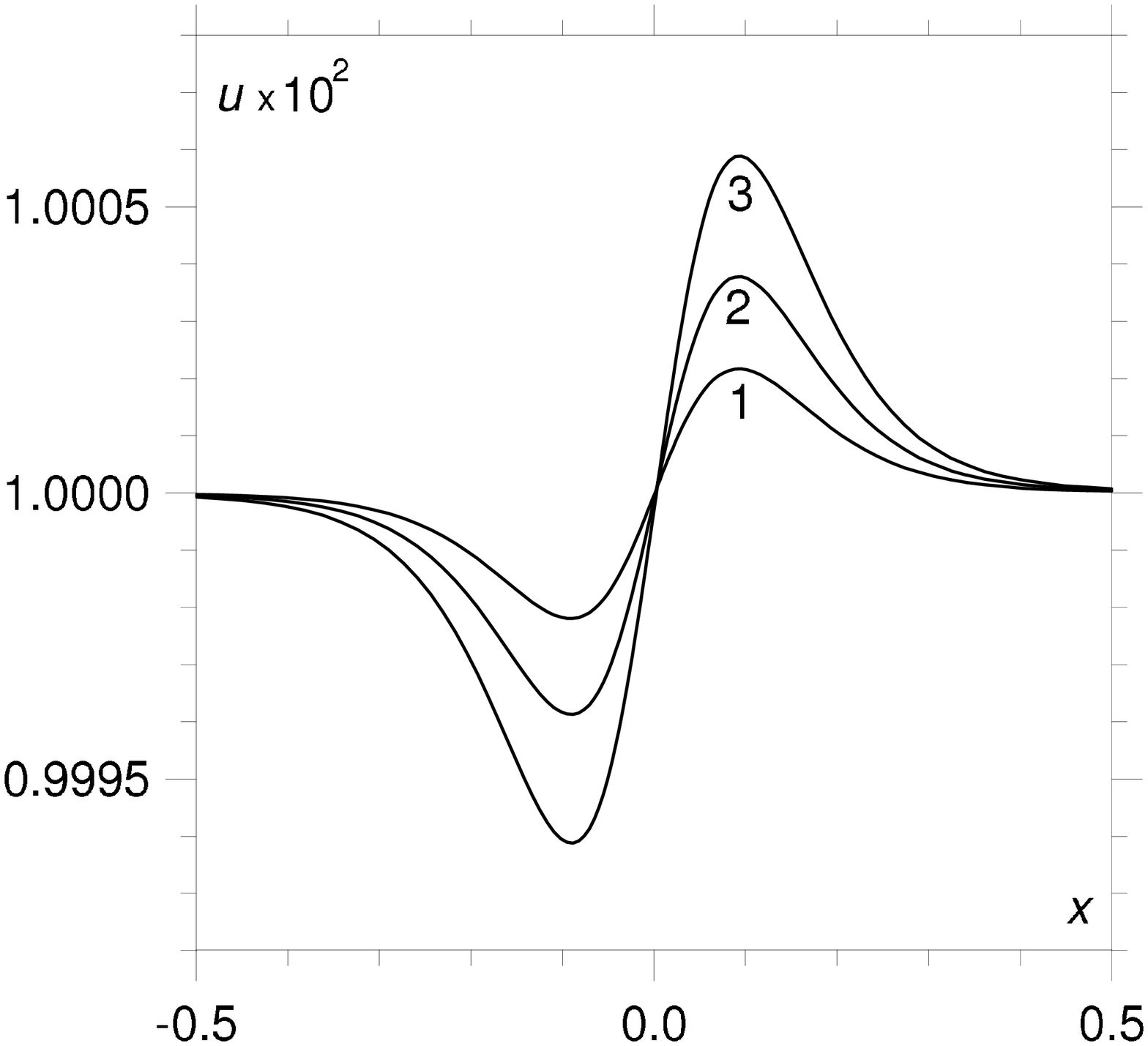}
\includegraphics[scale=0.3]{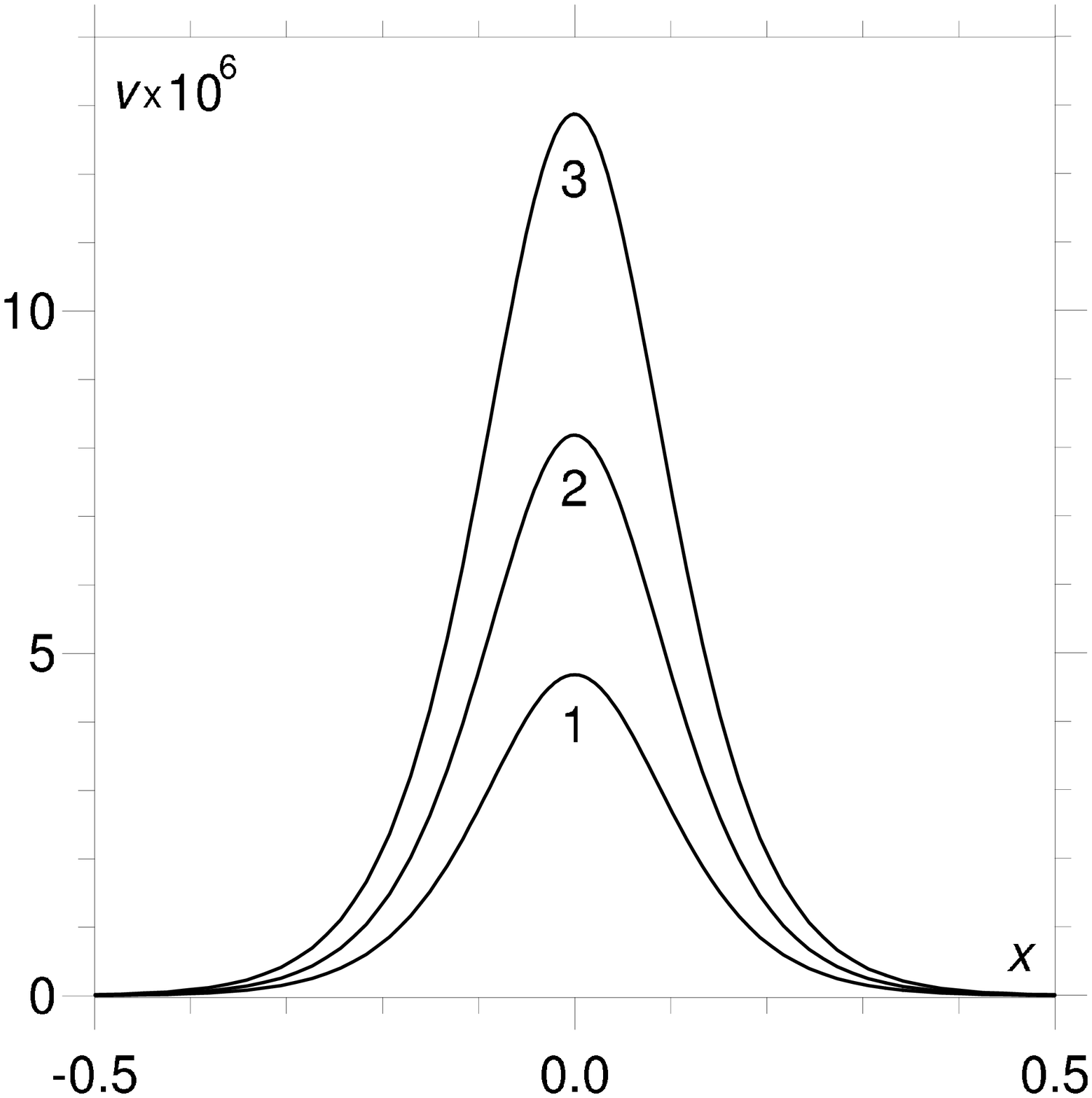}
\caption{Velocity on the surface for different solid-solid pairs.
Curve 1: $\theta_{1}=10^{\circ}$, $\theta_{2}=60^{\circ}$; 2:
$\theta_{1}=60^{\circ}$, $\theta_{2}=110^{\circ}$; 3:
$\theta_{1}=10^{\circ}$, $\theta_{2}=110^{\circ}$, respectively.
 For all curves $Re=0.01$, $\epsilon=0.01$, $\bar{\beta}=100$,
$Q=1$, $\bar{l}=0.1$, $\lambda=20$.} \label{F:wett}
\end{figure}

In Figs.~\ref{F:gen nav}--\ref{F:wett} the distributions of the
bulk velocity components along the solid surface are shown for the
case in which solid $1$ is more hydrophillic than solid $2$. In
each graph one can see (i) a positive normal velocity and (ii)
variation in tangential velocity on the surface.  The origin of
the first of these effects is relatively straightforward.  When
fluid particles forming the interface are driven by the outer flow
toward the region of lower equilibrium surface density, one has
that in the disturbed equilibrium $\rho^{s}>\rho^{s}_{e}$, and
hence, according to (\ref{norm fluxnd}), $v>0$.

Importantly, it can be seen in Fig.~\ref{F:stream} that the flux
out of the surface phase occurs both in the hydrophillic $(x<0)$
and hydrophobic $(x>0)$ regions of the solid and extends itself
well outside the transition zone. When solid $1$ is more
hydrophobic than solid $2$ one observes the reverse effect, with
the normal component of velocity directed towards the surface,
corresponding to a flux into the surface phase.  Once again this
occurs on both sides of the transition region.

In order to understand the distribution of the tangential
component of the bulk velocity on the surface we consider the
terms on the left-hand side of the generalised Navier condition
(\ref{gen navnd}) which determine $u(x,0)$. For a plane-parallel
shear flow over a homogeneous surface, i.e.\ far away from the
wettability transition region, there is no deviation of the
surface density from its equilibrium value $(\rho^s=\rho^s_e)$ and
hence no desorption from (or adsorption into) the liquid-solid
interface $(v=0)$. This is an obvious solution of
(\ref{incomnd})--(\ref{far fieldnd}). In dimensionless variables,
the tangential stress on the surface, $\partial u/\partial y +
\partial v/\partial x$, has magnitude $1$ ($\partial u/\partial y=1$,
$\partial v/\partial x=0$).

As one approaches the transition region, the situation changes.
Curve $1$ in Fig.~\ref{F:gen nav} shows the distribution of the
deviation of the tangential stress (the first term on the
left-hand side in (\ref{gen navnd})) from the value 1 across this
region. Importantly, there are now contributions to the tangential
stress from both $\partial u/\partial y$ and $\partial v/\partial
x$ since the latter becomes nonzero due to variation in the normal
velocity along the surface as a result of the spatially nonuniform
desorption (adsorption) process caused by the deviation of
$\rho^s$ from its local equilibrium value. As discussed earlier,
this effect could follow neither from the standard Navier
condition nor from any of its generalizations if the interface
formation process and the associated mass exchange between the
interface and the bulk are not taken into account.

The other factors that, according to (\ref{gen navnd}), lead to
slip on the liquid-solid interface behave as follows. As mentioned
in the previous section, in equilibrium the surface force,
$\left[\left(\lambda\rho^{s}/\rho^{s}_{e}\right) d \rho^{s}_{e}/d
x\right]/2$, and the surface tension gradient, $\left(d \sigma/d
x\right)/2$, balance each other to ensure that there is no
perpetual motion. However, when a shear flow in the far field is
imposed, it disturbs the surface phase, and the above two terms
are no longer in balance since they obviously depend differently
on the variable surface density $\rho^s$.  The deviation of the
second term on the left-hand side of (\ref{gen navnd}) from zero
is shown as curve $2$ in Fig.~\ref{F:gen nav}. One can see that it
is the variation in this second term in (\ref{gen navnd}) that
dominates and hence it is the imbalance between the surface
tension gradient and the tangential surface force caused by shear
flow that is mainly responsible for a variation in slip on the
surface as its wettability changes. The corresponding right-hand
side of (\ref{gen navnd}) and the normal velocity for these
parameter values are shown by curve $3$ of Fig.~\ref{F:eps_beta}.

Fig.~\ref{F:eps_beta} illustrates the effect that the size of the
parameters $\epsilon$ and $\bar{\beta}$ have on the solution. As
was mentioned, all effects associated with deviations from
no-slip, most clearly the nonzero normal velocity, are
proportional to $\epsilon$ and $\bar{\beta}^{-1}$, and the results
shown in the figure support this conclusion. Noticeably, the
normal velocity on the surface appears to be symmetric about the
centre of the transition region, where it achieves its maximum
value. The tangential velocity seems to be antisymmetric about the
origin, with a decrease in slip on solid $1$ and increase in solid
$2$.

Fig.~\ref{F:re_q_la} shows the effect on the solution of the
parameters $Re$, $Q$ and $\lambda$. Curves $2$ and $3$, which are
almost indistinguishable, demonstrate that the Reynolds number has
almost no effect on the dynamics of the surface. The {\it local\/}
Reynolds number near the surface is always small due to the
smallness of velocities in the vicinity of the solid surface even
when the value associated with the global flow is relatively
large, and one should indeed expect inertial effects to have
little impact on dynamics in the surface phase.

The parameter $Q$ determines the degree of mass exchange between
bulk and surface phase (\ref{norm fluxnd}). Curves $2$ and $4$ in
Fig.~\ref{F:re_q_la} show that the greater its value, the more
pronounced the effect will be. By varying $\lambda$ one can see
that the effect increases with the compressibility of the fluid.

Fig.~\ref{F:l} shows that for a smaller transition region one has
a sharper effect. In all cases the disturbance, i.e. the non-zero
normal component of the bulk velocity and variation in slip, runs
well outside the region of varying wettability. The results
suggest that if instead of a region, one treated the transition in
wettability as a solid-solid-liquid contact line, then the
disturbance to the bulk flow would be sharper than the effect
associated with a finite transition region but would still occur
over a finite region of the interface.

It is interesting to note that although the shape of the curves in
Fig.~\ref{F:l} differ, the integral of the normal velocity, i.e.
the total flux out of the surface phase per unit time
\begin{equation}\label{nf} J = \int^{\infty}_{-\infty} v~
dx ,\end{equation}
which we consider as a measure of the effect that a patterned
surface has on an adjacent flow, remains unchanged.  The value of
$J$ is a sensible choice of measure as we have seen that it is the
normal component of velocity that causes the noticeable deviation
from plane-parallel shear flow.

\begin{figure}
\centering
\includegraphics[scale=0.3]{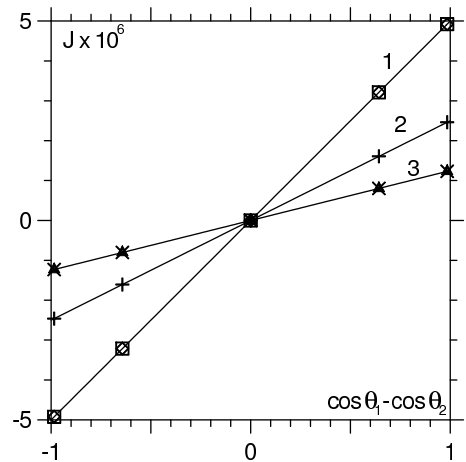}
\caption{Dependence of flux on parameters $Q$, $\epsilon$,
$\bar{\beta}$ and $\lambda$.  Parameters are varied around a base
state, the results of which are represented by vertical crosses,
of: $\bar{\beta}=100$, $Q=1$, $\epsilon=0.01$, $\lambda=20$. Then,
diamonds: $\epsilon=0.02$; squares: $Q=2$; triangles:
$\bar{\beta}=200$; diagonal crosses: $\lambda=40$.  For all curves
$\theta_{1}=90^{\circ}$ whilst $\theta_{2}$ is varied, and
$Re=0.01$ . Curves $1-3$ represent the predicted flux given by
$(21)$.} \label{F:numexp}
\end{figure}

Given that the change in solid does indeed alter the flow of an
adjacent liquid, consider how the magnitude of the effect is
dependent on the choice of solids. This can be illustrated by
looking at three different solids characterised by contact angles
of $10^{\circ}$, $60^{\circ}$ and $110^{\circ}$ that a free
surface would form with them.  Fig.~\ref{F:wett} shows the
velocity components for three different combinations of solids for
the case where the first solid is more hydrophillic.
The results suggest that the normal flux per unit time is
proportional to the difference $\cos\theta_{1}-\cos\theta_{2}$
which explains the ordering in Fig.~\ref{F:wett}, where we can see
the case $10^{\circ}\rightarrow60^{\circ}$ gives a far smaller
effect than $60^{\circ}\rightarrow110^{\circ}$ despite the
difference in contact angles being the same. The numerical
analysis of the problem made it possible to advance and then
verify the following approximate formula for $J$:
\begin{equation}\label{eqn} J =\frac{\epsilon Q }{2\bar{\beta}\lambda}
 \left(\cos\theta_{1} -
\cos\theta_{2}\right).
\end{equation}
This formula, which very accurately represent the numerical data
over a wide range of parameter values, has been obtained as
follows. The proportionality of the flux between the surface and
bulk phases to the cosines of the contact angle is in fact what
one should expect given that according to (\ref{state}) and
(\ref{young}), $\cos\theta$ is a linear function of the
equilibrium surface density, and it is the difference
$\bar{\rho}^{s}_{1e}-\bar{\rho}^{s}_{2e}$ that determines the
flux. This conjecture of the linear dependence of $J$ on
$\cos\theta_1-\cos\theta_2$ is the only nontrivial step in
obtaining (\ref{eqn}) and it has to be verified numerically. We
also have to find the coefficient of proportionality in this
dependence in terms of the dimensionless parameters of the
problem. In order to do this, we used the following procedure. For
the fixed values of $\theta_1$ and $\theta_2$, we chose a base
state in terms of the remaining parameters and then varied one of
these parameters away from its base state to consider the effect
this variation produces on $J$. After repeating this operation for
all parameters we arrive at equation (\ref{eqn}) which now has to
be verified by independently varying all dimensionless parameters,
including $\theta_1$ and $\theta_2$. The accuracy with which this
equation represents $J$ is illustrated in Fig.~\ref{F:numexp}.

Order of magnitude arguments for the phenomenological parameters
and the analysis of experiments on dynamic wetting \cite{blake02}
suggest that $\beta\sim \mu/h$, where $h$ is the thickness of the
interfacial layer (modelled here as an `interface' of zero
thickness). Using this estimate, for the dimensional flux per unit
time out of the surface phase in a liquid/solid/solid system one
has
\begin{equation}\label{eqnd}
J_{dim} \sim \frac{S h \sigma_{lg}}{\rho\gamma}
\left(\cos\theta_{1}-\cos\theta_{2}\right).
\end{equation}
Given that $h$ is typically very small (a few nm for simple fluids
\cite{rowlinson82}), the above estimate highlights the subtle
nature of the effects that we have described.

\section{Conclusion}

As was shown, the interface formation model applied to the flow
over a solid surface of variable wettability allows one to
describe the main features of this flow observed in molecular
dynamics simulations, most notably the nonzero component of the
bulk velocity normal to the solid surface.  A natural link between
`wettability' interpreted in terms of the concept of the `contact
angle' featuring in the spreading of liquids on solid surfaces and
a viscous flow over chemically-patterned solids with no free
surface present has been established. Importantly, the interface
formation model deals with these phenomena entirely within the
approach of continuum mechanics with no artificial inclusion of
intermolecular forces in its framework. An interesting feature
that follows from the results is that slip, i.e.\ the difference
between the tangential component of the fluid's velocity and the
corresponding component of the velocity of the solid surface,
results primarily from the disturbance of the force balance in the
`surface phase' and not from the tangential stress, as follows
from the standard Navier condition.

\begin{acknowledgments}
The authors kindly acknowledge the financial support of Kodak
European Research and the EPSRC via a Mathematics CASE award.
\end{acknowledgments}

\bibliography{manuscript} 

\end{document}